\documentclass[aps,pre,reprint,groupedaddress,doi=false,isbn=false,url=false,title=true,longbibliography]{revtex4-2}
\usepackage{amsmath}
\usepackage{amssymb}
\usepackage{float}
\usepackage{graphicx}
\begin{document}
\title{Nonflat Histogram Techniques for Spin Glasses}
\author{Fabio M\"uller}
\email{fmueller@itp.uni-leipzig.de}
\author{Stefan Schnabel}
\email{schnabel@itp.uni-leipzig.de}
\author{Wolfhard Janke}
\email{janke@itp.uni-leipzig.de}
\affiliation{Institut f\"ur Theoretische Physik, Universit\"at Leipzig,
  IPF 231101, 04081 Leipzig, Germany}
\date{\today}

\begin{abstract} We study the bimodal Edwards-Anderson spin glass comparing
established methods, namely the multicanonical method, the $1/k$-ensemble and
parallel tempering, to an approach where the ensemble is modified
by simulating power-law-shaped histograms in energy instead of flat
histograms as in the standard multicanonical case.  We show that by this
modification a significant speed-up in terms of mean round-trip times
can be achieved for all lattice sizes taken into consideration.
\end{abstract}
\newcommand{\frechet}{Fr\'{e}chet\ }
\newcommand{\epsmean}[1]{\langle \tau\rangle_{#1}}
\newcommand{\SH}{{P}_{\text{SH}}}
\maketitle
\section{Introduction}
Simulations of systems with rugged free-energy landscape~\cite{Janke2008a}
suffer from massive slowing down of the dynamics in the low-temperature phase.
This problem encountered in many physical systems, e.g., folding
polymers or spin glasses, renders the investigation of the thermodynamical
properties of such systems in the low-temperature phase a very challenging
task.

The Metropolis algorithm~\cite{Metropolis1953} is designed to sample configurations according to their statistical weight in the canonical ensemble.
At low temperatures it fails dramatically for systems with rugged free-energy
landscape because of the effectively broken ergodicity. The simulations
get stuck in local minima (metastable states) and the thermal energy is not sufficient
to overcome the huge free-energy barriers.  There have been a wide range of
algorithmic developments tackling this problem that can all be subsumed in the
term of broad-energy ensembles.

One commonly employed method is the
Parallel Tempering (PT)~\cite{Hukushima1996,Swendsen1986,Geyer1995} method where
Metropolis simulations of copies of the system (replicas) at different temperatures are
performed.  After certain time intervals exchanges of the replicas between the
different temperatures are attempted.  This
procedure enables the replicas at low temperature to fully explore deep
free-energy valleys and at high temperature to travel freely through the phase space
and thus to decorrelate.  Thereby, the different replicas can explore the rugged
structure of the free-energy landscape much more efficiently than in a simple
Metropolis simulation.  Using this method at temperatures close to the transition,
studies of spin-glass systems of sizes up to $40^3$ spins have been reported~\cite{Hasenbusch2008,Baity-Jesi2013}. For
ground-state searches systems of about $10^3$ are feasible~\cite{Wang2015a}. The great
advantage of PT is its
simplicity: the algorithm only needs a suitable temperature set and exhibits
good performance which makes it probably the most employed method in the
investigation of systems with rugged free-energy landscape.

Another recent development is the Population Annealing Monte
Carlo~\cite{Iba2001,Hukushima2003,Machta2010,Wang2015,Barash2017a} method which proceeds similarly to
Simulated Annealing~\cite{Kirkpatrick1983} as the system is gradually cooled down according to an
annealing schedule.   The annealing is performed on a big population of replicas and by introducing intermediate resampling of the population of replicas after lowering the temperature the simulation is kept at thermal equilibrium.  This permits the evaluation of thermodynamic observables in contrast to simple Simulated Annealing.  Despite the
attempts of optimizing the method for spin
glasses~\cite{Hukushima2003,Wang2015,Barzegar2018} it is not able to outperform PT.  Its optimization, however, remains more cumbersome due to the additional
complexity.  The main advantage of this algorithm is its suitability for
massively parallel implementation.  For disordered systems, however, this advantage does
not come into play, because the necessity of simulating many different disorder
realizations allows the efficient use of parallel computing for any
method.

The Multicanonical (MUCA) method~\cite{Berg1991,Berg1992,Janke1992} is another well established algorithm
designed for the simulation of systems with first-order phase transitions which
performs well in the simulation of systems with rugged free-energy
landscape, too.  It has already been applied to spin glasses in Refs.~\cite{Berg1992a,Berg1993}.
In this method the simulation is set up to visit all possible energies
with the same probability yielding a flat histogram in energy.
However, it has been noted by different researchers that this ensemble is not optimal.
One suggested improvement is the $1/k$-ensemble by Hesselbo and
Stinchcombe~\cite{Hesselbo1995}, where the sampling distribution is the inverse
of the integrated density of states.  As the authors point out, this description
samples the low-energy region more often than the high-energy region, resulting in
energy histograms which grow towards the low-temperature phase.

Another non-parametric optimization of the MUCA algorithm was proposed in
Ref.~\cite{Trebst2004}.  The method uses an estimator of the local diffusivity
in order to maximize the number of performed round trips in energy.  The method
is among others applied to the ferromagnetic Ising model for which it
improves the scaling behavior of the round-trip times in energy.
The improved performance
for the models considered in that work and the nature of the algorithm of
automatically identifying the bottlenecks of the simulation and
concentrating the simulation effort on
this region suggest that the round-trip times of the simulations should
diminish independently of the underlying system.
However, in our implementation, in the case of the three-dimensional (3D) bimodal
Edwards-Anderson (EA) spin glass~\cite{Edwards1975},
the round-trip times did not systematically improve with this method.
Instead the
simulation got stuck for some of the considered samples, rendering a comparison
to the other methods impossible.

In this work we present a different approach:
we prescribe parametric profiles for the histograms of the simulation and adjust
the simulation weights accordingly.
As for the three previous MUCA variants, it requires the
knowledge of the underlying density of states, but it is
much more flexible.  The profiles are all chosen to be shifted power laws having
two free parameters.

As an example we consider the 3D bimodal EA spin glass.  This is one of the simplest models
exhibiting a rugged free-energy landscape and is also interesting from
the point of view of an optimization problem where finding ground states of hard
disorder realizations is NP-hard~\cite{Barahona1982}.  Despite the exponential
growth of the computational resources fundamental questions regarding the
nature of the spin-glass phase still remain.
For the progress in understanding the open questions the development
of new methods and an improvement of the existing methods is crucial.

The rest of the paper is organized as follows.  In
Sec.~\ref{sec:Model} the spin-glass model and the simulation methods are
explained.
The direct comparison of the round-trip times of the individual
methods is performed in Sec.~\ref{sec:scattering}.
The framework of extreme-value statistics is introduced in
Sec.~\ref{sec:distributions}.
In Sec.~\ref{sec:assessing_performance} benchmarks for the global comparison
are discussed and the different methods are compared in terms of those
benchmarks.
The results are summarized in Sec.~\ref{sec:conclusion}.

\section{Model and Employed Methods}
\label{sec:Model}

We take into consideration the 3D bimodal EA model whose
Hamiltonian takes the form
\begin{equation}
      \label{eq:hamiltonian}
    H = -\sum_{\langle ij\rangle}J_{ij}S_{i}S_{j},
  \end{equation} where the bonds $J_{ij}$ and the spins $S_i$ can take values
$\pm 1$.  The sum runs over all neighboring spins in the simple-cubic
lattice with periodic boundary conditions.

Due to the disordered nature of spin glasses the
study has to take into account a sufficiently large set of disorder realizations
on which the averaged quantities can be computed.  In this case one disorder
realization consists of a set of $3V$ couplings $J_{ij}$ which are either
positive or negative unity with a probability of 50\%, where $V=L^3$ is number of
spins in a lattice of linear lattice size $L$. The disorder realizations are
generated prior to the simulation and then kept fixed for all times (quenched
disorder).  As an adequate set of disorder realizations 4000 samples with $L=3$
and $L=4$ are generated and 5000, 6000, and 4000 samples of size $L=5,6$, and $8$,
respectively.

The method which we adapted is the well-established MUCA
method~\cite{Berg1992} employing a generalized Metropolis criterion with an
energy dependent weight function
\begin{equation}
  \label{eq:muca}
  P_{\text{acc}}=\min\left( 1,\frac{W(E_\text{new})}{W(E_\text{old})}\right),
\end{equation} where the weight function is proportional to the inverse of the
density of states $\Omega(E)$,
\begin{equation}
  W(E)\propto \Omega^{-1}(E).
\label{eq:muca-weights}
\end{equation} For the MUCA simulations $\Omega(E)$ has to be sufficiently
well-known a priori for each disorder realization.  An estimator for it can,
for instance, be obtained by means of the Wang-Landau
algorithm~\cite{Wang2001} or, as in this work, by
other iterative procedures which are explained, e.g., in Ref.~\cite{Janke2003}.
This ensemble produces histograms which are flat in energy and is, therefore,
often also referred to as ``flat histogram method''.

A straightforward
generalization of the flat histogram method are the nonflat histogram methods.
If the simulation weights for the flat MUCA method are multiplied with the
desired energy dependent shape (or profile) function $\SH(E)$
\begin{equation}
   W(E)\propto \Omega^{-1}(E)\SH(E),
 \end{equation} the resulting histograms will be shaped according to $\SH(E)$.
In this work all the profiles are shifted power laws of the form
\begin{equation}
  \label{eq:power_law}
  \SH(E,\Delta E,\alpha)=\left(\frac{E}{\Delta E-E_g}+1\right)^{\alpha},
\end{equation} where the exponent $\alpha < 0$ and $\Delta E>0$ is the position of the pole relative to the
ground-state energy $E_g$ of the respective spin-glass realization.
\begin{figure}
  \centering
  \includegraphics{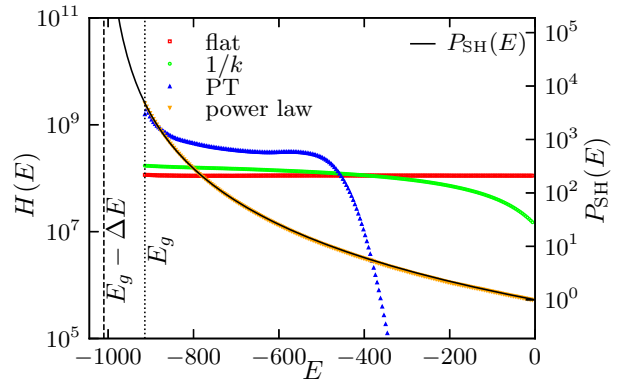}
  \caption{The recorded histograms $H(E)$ of the different methods and the profile function $P_\textrm{SH}(E)$
    for one disorder realization of linear lattice size $L=8$. The dotted
    and the dashed vertical lines indicate the position of the ground-state energy $E_g$ and the position of the
  pole of the power law~\eqref{eq:power_law}, respectively.}
  
  \label{fig:histos}
\end{figure}
In this parametrization
the power laws are normalized to unity at $E=0$.

In Fig.~\ref{fig:histos} the recorded histograms of the different methods are displayed
on a logarithmic $y$-scale for one disorder realization with $L=8$.  In contrast to flat MUCA  all methods
have in common that the distribution of sampled states grows towards the ground-state energy.
The recorded histogram of nonflat MUCA matches perfectly the imposed profile and its
histogram in the ground-state region is similar to that of PT.
We are convinced that this feature which among the existing methods is strongest for PT enhances
the ability of sampling the low-energy region and especially the ability of finding low-energy
states of investigated systems.  There are different possible choices of functional forms
which enhance the sampling of the low-energy region and even stepwise defined function could
be employed and might even yield better results.  We chose a power law because
the two involved parameters allow for a good adaptation but the tuning of the parameters in
the two-dimensional parameter space remains feasible. 

For the above parametrization we found
a fixed parameter set namely $\alpha=-3.6$ and $\Delta E=96$ which independently
of the lattice size yielded the shortest mean round-trip times, among the
considered profiles.  Subsequently we will refer to the nonflat MUCA setting
with the power-law shape belonging to this parameter set just as power-law (PL)
setting or nonflat MUCA method.  While the overall best results are obtained with this
parameter set, we want to point out that an improvement compared to flat MUCA was
visible for each of the considered parameter sets.  The parametrization with a fixed
offset from the ground-state energy yields different relative distributions depending
on the ground-state energy encountered in the respective disorder realization. The
value of the profile function at the ground-state energy is given by
\begin{equation}
  \label{eq:profile_ratio}
  \SH(E_g,\Delta E, \alpha)=\left(\frac{1}{1-\frac{E_g}{\Delta E}}\right)^\alpha.
\end{equation}
The sampling at the ground-state energy compared to zero energy is thus
enhanced by a factor of $\approx 13$ for a disorder realization with $L=4$ and a
typical ground-state energy of $\approx -100$.  For a sample with $L=8$ and typical
ground-state energy of $\approx -900$ instead it is enhanced by a factor of $\approx 4500$.
Due to this feature this parametrization of the profile function does not require any
adjustments of the parameters in the system sizes which we considered.  Presumably such a
profile will also yield good results for larger systems, although we cannot be certain.

Next the $1/k$-ensemble~\cite{Hesselbo1995} is considered which is defined by setting the
simulation weights equal to the inverse of the integrated density of states up to the energy of
the respective bin
\begin{equation}
  \label{eq:1/k}
  W_{1/k}(E)\propto 1/k=\left(\int_{E_g}^E dE'\Omega(E')\right)^{-1}.
\end{equation}
Here, a first-order Taylor expansion of $\ln\Omega$ at $E$ leads to
$W(E)\approx W_{1/k}(E)$ if $P_{\rm SH}(E)=d\ln\Omega(E')/dE'|_{E'=E}$.  This prescription again relies on the knowledge of the density of
states.  The authors of Ref.~\cite{Hesselbo1995} stress its robust ergodicity and apply it to spin glasses
and the traveling salesman problem~\footnote{The aim of the study was not
  maximizing the number of round trips in energy but rather the amount of
  statistically independent data in an uncorrelated Monte Carlo
  simulation.}.

\begin{figure*}
  \centering
\includegraphics{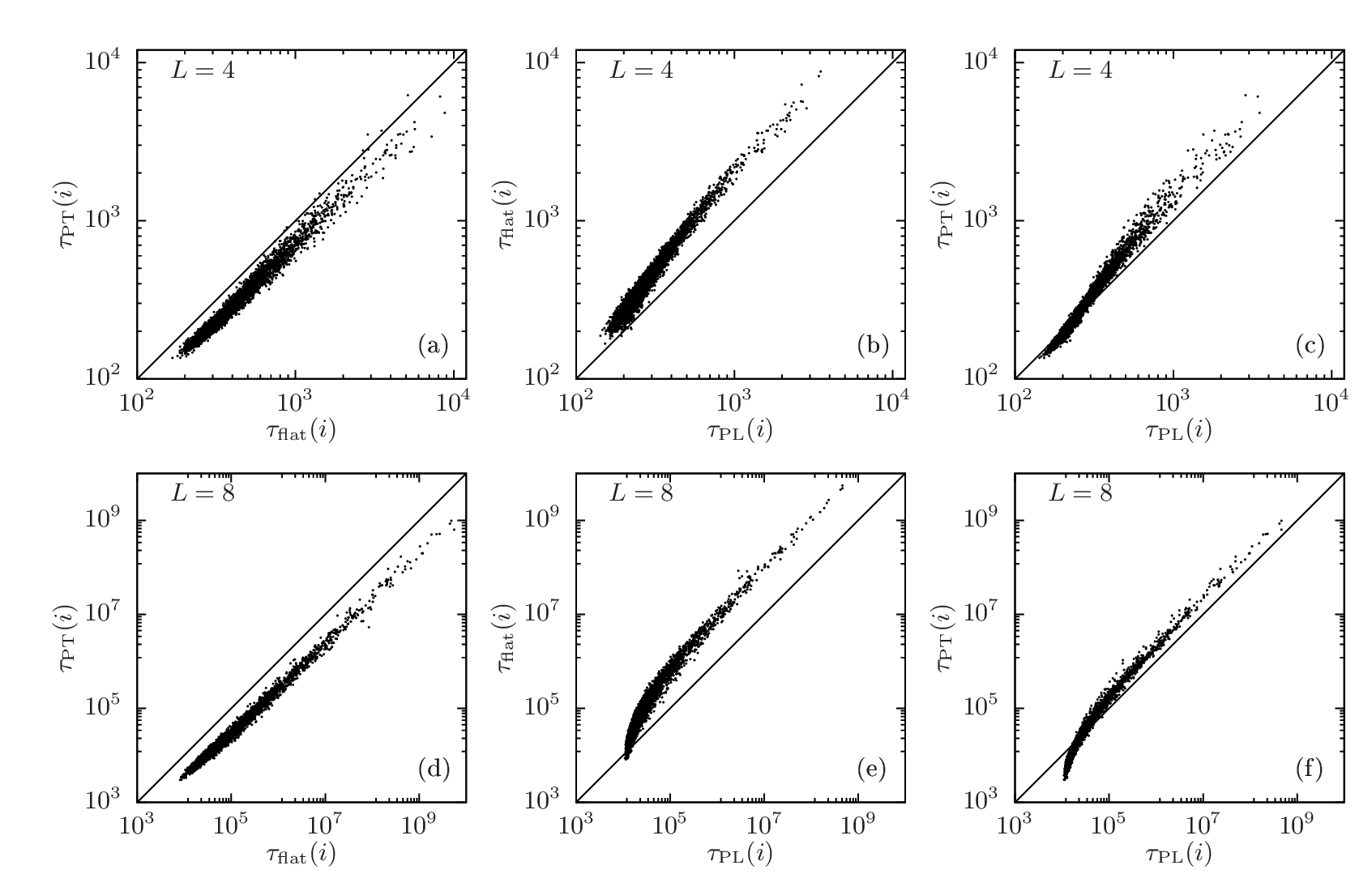} 
  \caption{Scatter plots of the round-trip times comparing the nonflat power-law histogram technique (PL) to the standard flat MUCA (flat)
and parallel tempering (PT) methods for sizes $L = 4$ (upper panels) and $L = 8$ (lower panels). All points scattered above the identity line have
longer round-trip times for the method on the $y$ axis.}
  \label{fig:scattering_plots}
\end{figure*}
Since for the above mentioned methods the density of states is the only needed input
it was determined only once
to high accuracy employing the iterative procedure adapted from Ref.~\cite{Janke2003}
but with power-law shaped distributions in energy.
In this case, and generally when the ground-state energy of the system is not known,
a priori the profile function has to be adapted whenever a lower energy is found.

Lastly, the PT method being
probably the most employed algorithm for spin-glass simulations, is included
in the comparison.  The ensemble in this case is defined by a set of $M$
temperatures $\{T_i,\,i=1,...,M\}$. For each temperature $T_i$ a Metropolis simulation of a
copy of the system (replica) is performed.  The temperatures of the replicas $i$
and $j$ are allowed to exchange configuration according to
\begin{equation}
  \label{eq:temperatureexchange}
P_{ij}^\mathrm{ex}=\min\left( 1,e^{(\frac{1}{T_j}-\frac{1}{T_i})(E_j-E_i)}\right),
\end{equation}
where $E_i$ and $E_j$ are the energies of replica $i$ and $j$ and $k_B=1$. This
prescription allows for fast decorrelation when a replica travels to high
temperature and the exploration of the local minima at low temperatures.  Among
the vast choice of different PT protocols
available~\cite{Papakonstantinou2014} we opted for the constant exchange rate
protocol with acceptance rates between 40\% and 60\%~\cite{Bittner2008}.  For all simulations
the maximal temperature was chosen to be well above the critical temperature,
$T_\mathrm{max}>3>T_c\approx 1$.
The exchange rates were
imposed on each individual disorder realization in an initial equilibration run
during which the temperatures were modified accordingly.
The number of replicas was set to $M=7,7,12,14, \text{and } 20$
for $L=3,4,5,6, \text{and } 8$, respectively.  We note
that the choice of the temperature set is crucial for the PT algorithm and also
provides the possibility of optimizations as for example in Ref.~\cite{Katzgraber2006}.
However, in this work we rather limit ourselves to a well established protocol
for PT focusing on the
optimization of the nonflat histogram technique.

\section{Comparison of the Round-Trip Times}
\label{sec:scattering}
The observable taken into account for this study is the round-trip time.
For all methods except PT and each disorder realization it is defined as the time needed by the simulation to travel from the highest energy ($E\approx 0$) to the ground-state energy and back.  For PT, instead, the round trip is measured between the ground-state energy and an energy typical for a canonical ensemble 
with a temperature well above the freezing point of the disorder
realization~{\footnote{The estimated temperature was extracted from few single disorder
  realizations for each lattice size and then taken as constant for all samples
  with the same lattice size.}\footnote{This different measuring prescription gives PT a slight advantage
in comparison to the other methods which is, however, negligible in the authors' opinion.}}.
This time can be taken as an upper bound of the autocorrelation time of the energy of
the respective disorder realization at the ground state.  We want to stress that
the energies we refer to as ground-state energies are the lowest encountered energies
and may not be the true ground states.  However, the round-trip times were
always measured performing at least 100 round trips for each individual sample and method
so that several hundred round trips have been performed on each disorder realization.
In case lower energies were measured during this process the disorder realization
was requeued and simulated again until the desired number of round trips
was achieved.  This procedure renders the discovery of the true ground state
very probable.  After at least 100 round trips the relative statistical error
in the round-trip time $\tau_i$ is of the order of $\Delta \tau_i/ \tau_i \approx 0.1$.
\begin{figure*} \includegraphics{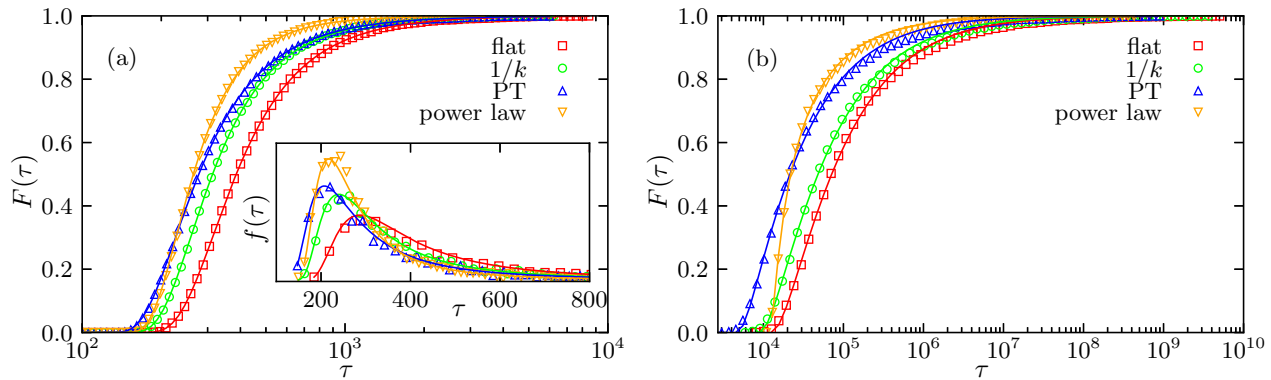}
  \caption{Round-trip time distributions (symbols) and best fitting cumulative
distribution functions (lines) for the different methods and lattice size $L=4$
(a) and $L=8$ (b).  The inset of the left panel shows the PDF form of the distribution.}
  \label{fig:cdfs}
\end{figure*}

The first property we want to look at is the dependence of the round-trip times
for the individual disorder realizations on the employed method.
The scatter plots in
Fig.~\ref{fig:scattering_plots} show the round-trip times of the same disorder
realization for two different methods for all the simulated disorder realizations
of size $L=4$ and $L=8$ on a log-log scale.   The strong
correlation of the round-trip times for each single disorder realization should
be noted, indicating that the hardness of the underlying optimization problem is
primarily a characteristic of the disorder realization and mostly independent of the
employed method.  This fact allows us to categorize the disorder realizations
and speak of easy and hard instances.
Comparing the round-trip times $\tau_i$ for the flat MUCA
method and the parallel tempering method (left panels) for both lattice sizes
$L=4$ and $L=8$, the $\tau_i$ are systematically lower for PT,
indicating its superior performance for the whole classes of the
bimodal EA spin glasses of the respective lattice sizes.

When comparing the
performance of the nonflat histogram method to the flat MUCA method (central
panels) the surrounding area of the scattered round-trip times shows a bending,
i.e., for $L=4$ the flat histogram method displays only slightly higher round
trip times for the easy disorder realizations.  With increasing hardness the
round-trip times for the flat histogram method grow faster than for the
PL setting.  This effect gets enhanced with a further increase of the lattice size (see lower panel) where for the case of $L=8$ the round-trip times of the easiest samples for the flat MUCA method are similar to those for PL.
However, as will become apparent in the next section, the hard
samples contribute most to the mean round-trip time so that even a slightly
weaker performance for the easier samples would hardly contribute to the total computation time.

The right panels show the comparison of PL to PT.  For $L=4$
PT outperforms the nonflat histogram method for the
easy disorder realizations, while for the hard ones PL
displays shorter round-trip times.  For $L=8$ a large fraction of the disorder
realizations is characterized by shorter round-trip times for PT, but
the tail of the distribution describing the hard samples
exhibits shorter round-trip times for PL.

\section{Round-Trip Time Distributions}
\label{sec:distributions}

In order to quantify the observations of the previous section the distributions
of the round-trip times can be examined.  
Round trips in energy include the visit
of the ground state of the respective disorder realization which is an extreme
event.  Their statistics must thus be described in the framework of
extreme-value statistics.  One of the main results in this field is given by the
Fisher-Tippet-Gnedenko
theorem~\cite{Charras-Garrido2013} which characterizes the type of distributions which
extreme-value distributions can converge to.  The round-trip time distributions
of the bimodal EA spin glass all seem to converge to \frechet
distributions independently of the method and the system size.  This has
already been suggested in Ref.~\cite{Alder2004} for the round-trip time
distributions of the 3D EA model employing the flat histogram ensemble.

One parametrization of the cumulative distribution function (CDF) of the \frechet
distribution is given by
\begin{equation}
  \label{eq:cdf}
F(\tau)=\exp\left[-\left(1+\xi\frac{\tau-\mu}{\beta}\right)^{-1/\xi}\right]
\vphantom{\exp\left(-\left(1+\xi\frac{\tau-\mu}{\beta}\right)^{-1/\xi}\right)
_{\frac{a}{A}}^{\frac{1}{\tau}}},
\end{equation} with $\tau \in [\mu-\beta/\xi,\infty)$.  The location of the
distribution along the $\tau$-axis is determined by $\mu$, $\beta$ is the scale parameter, and
the shape parameter $\xi$ describes the decay of the tail of the
distribution, i.e., the occurrence of rare events.  The CDF is
the integrated form of the probability density function (PDF) $f(\tau)$.  The round-trip time
distributions are thus all defined by sets of parameters $\mu,\beta,\xi$ which are determined by fitting the CDF to the recorded round-trip times.
\begin{figure}
  \centering

\includegraphics{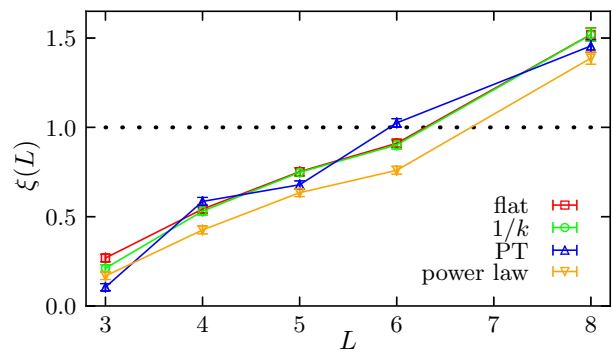}
  \caption{The figure shows the shape parameter of the best fitting \frechet
distribution in dependence of the lattice size for the different employed
simulation methods.  The dotted line indicates the threshold value from which on
the distribution mean diverges.}
  \label{fig:scaling2}
\end{figure}

The measured round-trip times and the respectively best fitting \frechet
distribution for lattice sizes $L=4$ and $L=8$ are plotted in Fig.~\ref{fig:cdfs}.
The points represent the measured data and the solid lines are the best fitting
\frechet distributions.
The varying performance of the methods in dependence on the difficulty of the
disorder realizations which became visible in the last section,
also reflects in the distribution of the round-trip times.  For both lattice sizes the
CDF belonging to the flat MUCA method is lower for all $\tau$ than the one belonging
to PT.  The maximum increase which corresponds to the bulk of the distribution is
shifted to higher $\tau$ for MUCA as compared to PT.

Comparing PT instead to the PL setting yields a different
picture: the cumulative distribution functions for lattice size $L=4$ cross
at $F(\tau)\approx 1/3$, corresponding to a round-trip time
$\tau\approx 2\times 10^2$.  This means that for the PT algorithm
the easiest one third of all samples have smaller round-trip times than the
easiest third for the PL method, while PL is faster for the harder two thirds.
For $L=8$ the PL round-trip times are larger for the easier half
of the samples and smaller for the harder half.
The round-trip times
for the hard disorder realizations have most influence on the decay of the
distribution and thus on the shape parameter $\xi$.
In Fig.~\ref{fig:scaling2} the scaling
of the shape parameter for the different methods is displayed, where the errors
of the best fitting parameters are estimated via jackknifing~\cite{Efron1982}.
For the considered lattice sizes the shape parameter scales similarly for all
the different methods.  However the values for PL are
systematically lower for $L\geq 4$.  This is in good agreement with its superior
performance for the difficult disorder realizations.
\begin{figure}
  \centering
  \includegraphics{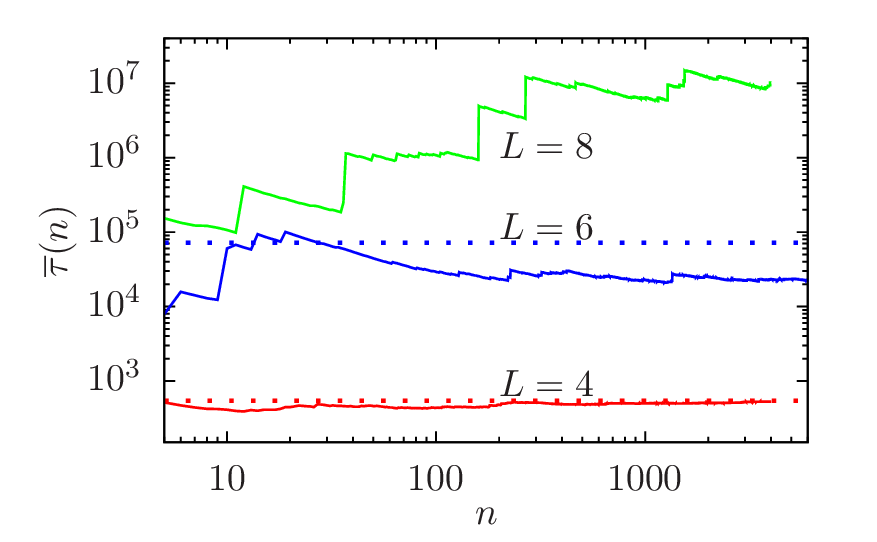} 
  \caption{Illustration of the convergence of the population mean of the
    round-trip times for the flat MUCA method in dependence of the
    population size to the distribution mean of the
underlying distribution.  The solid lines are the running mean including the first
$n$ samples and the dotted lines in the respective color are the means of the
underlying distribution.}
\label{fig:running_mean}
\end{figure}

\section{Assessing the Performance of the Different Methods}
\label{sec:assessing_performance}
Next, we want to compare the performance of the different simulation methods. The
most intuitive observable would be the disorder average of the round-trip times
over the set of considered disorder realizations.  However, as it will turn out
the rare-state events which have a dominating influence on the distribution mean
are not within the set of simulated disorder realizations.
This effect is accounted
for by considering distribution means up to large quantiles of the underlying
extreme-value distributions, yielding a more reliable measure of the \emph{real}
performance of the different methods.

\subsection{Finding a Benchmark}
In principle the \emph{real} performance could be determined by measuring the
round-trip time of every possible disorder
realization.  This procedure is discarded due to the enormous number of possible
disorder realizations~\footnote{The bimodal EA spin glass, having discrete
  randomness, has only a finite number of possible disorder realizations.  Due
  to the symmetries in the absence of external
  fields this number can be estimated to be of the order of $2^V$, where $V=L^3$ is the number of spins contained in the lattice. }.
Instead
we generate a subset of all possible disorder realizations and from those we try
to infer the expected mean round-trip time of all the disorder realizations
belonging to the same problem class, by means of the \emph{population mean} $\tau_{\rm pop}$.
This is a standard approach in all Monte Carlo
studies and the law of large numbers assures its convergence for all random
variables from distributions with well defined mean.
However, this prerequisite is not fulfilled for all of the round-time
distributions encountered in this work.

The expected mean round-trip time resulting from the underlying
probability density could be estimated by the \emph{distribution mean}
\begin{equation}
 \langle\tau\rangle=\int\limits_{\mu-\beta/\xi}^{\infty}d\tau  \tau f(\tau).\label{eq:int_mean}
\end{equation}
This integral can be computed analytically, yielding
\begin{equation}
  \label{eq:frechet_mean} \langle \tau\rangle=\begin{cases}
\mu+\frac{\beta}{\xi}\left[\Gamma \left(1-\xi\right)-1\right] &{\text{for }}\xi < 1 \\
\infty &{\text{otherwise}}\end{cases},
\end{equation}
with $\Gamma(x)$ being the gamma function. 
The distribution mean is, therefore, only defined as long as the shape
parameter $\xi$ is smaller than one~\footnote{Due to the finite
  number of possible disorder realizations the mean \emph{is} actually
  defined.  Its finite value is expected to be of the order of $10^{100}$,
  and can therefore in terms of computation time numerically not be distinguished
  from a real divergence.}.
To illustrate this difficulty one can consider the \emph{running mean} which is defined as the population mean over the first $n$ generated disorder realizations keeping them in a fixed order,
\begin{equation}
  \label{eq:running_mean}
\overline{\tau}(n)=\frac{1}{n}\sum_{i=1}^n \tau_i,
\end{equation}
implying $\tau_{\rm pop}=\overline{\tau}(N)$, where $N$ is the number of all simulated disorder realizations.
\begin{figure*}
  \centering

 \includegraphics{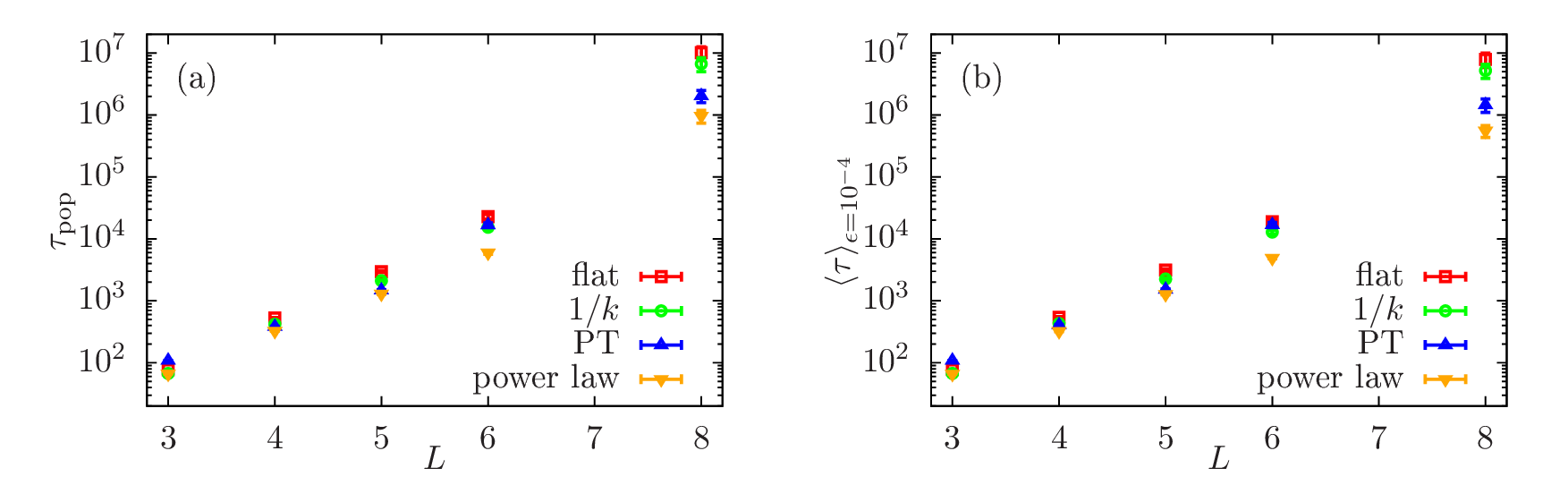}
 \caption{Population mean $\tau_\mathrm{pop}$ (a) and quantile mean $\langle \tau \rangle_{\epsilon=10^{-4}}$ (b) of the round-trip times for the different methods as a function of system size. The
 latter is the more reliable statistical quantity.}
\label{fig:means}
\end{figure*}

In Fig.~\ref{fig:running_mean} the running mean for the flat MUCA method and
different system sizes is plotted together with the respective distribution
mean, if it is defined.  For $L=4$
($\xi\ll1$) the running mean quickly converges to the distribution mean indicated
by the dotted line.  For $L=6$ ($\xi\approx 1$) the jumps due to rare events in
the tail of the distribution become more pronounced.  The running mean is still
expected to approach the distribution mean for a finite number of
disorder realizations.  For the 6000 samples considered in our work this is
still not the case.  For $L=8$ ($\xi> 1$) the distribution mean is not defined.
In the picture of the running mean, jumps represent round-trip times in the tail
of the distribution. In the case of $\xi> 1$ those jumps $\tau_n/n$ in the
running mean are clearly visible in Fig.~\ref{fig:running_mean}
and will lead to a divergence of the population mean the more disorder realizations
are taken into account and hence the more rigorously the tail of the
distribution is explored.  This illustrates that
the population mean as a measure for the performance of the different methods
must be taken with a grain of salt.

In order to retain the characteristics of the underlying round-trip time
distribution into the estimator of the performance of the different methods the
distribution mean up to a certain quantile can instead be taken into account.
The quantile function is the inverse of the CDF~\eqref{eq:cdf},
  \begin{equation}
    Q(p)=F^{-1}(p)=\mu+\frac{\beta}{\xi}\cdot\left[\left(-\ln p \right)^{-\xi}-1\right],\quad p\in(0,1),
\end{equation}
yielding the round-trip time $\tau_p=Q(p)$ at which a certain fraction $p$ of the
distribution is accumulated.
For each $\epsilon<1$ we define the \emph{quantile mean} $\langle \tau\rangle_{\epsilon}$ disregarding a fraction $\epsilon$ of the tail of the distribution as the integral~\eqref{eq:int_mean} with the upper bound replaced by $Q(1-\epsilon)$,
\begin{equation}
  \label{eq:quantile_mean}
  \langle\tau\rangle_\epsilon=\int\limits_{\mu-\beta/\xi}^{Q(1-\epsilon)}d\tau  \tau f(\tau).
\end{equation}
The integral is evaluated with the parameters of the best fitting distributions to
the measured round-trip times, see Fig.~\ref{fig:cdfs}.
This enables a well-defined extrapolation
beyond the measured round-trip times of the simulated disorder realizations of
the underlying study and thus a comparison of the different methods
beyond the mere population mean, which may be strongly dependent on the set of
disorder realizations taken into account for the study.                        
\subsection{Comparison of the Different Methods}
\label{sec:mean-round-trip}

Finally, for the comparison of the mean round-trip times only the population mean $\tau_{\rm pop}$ and the
quantile mean $\langle \tau \rangle_{\epsilon=10^{-4}}$ neglecting a fraction $\epsilon=10^{-4}$ of the tail of the
distribution are taken into account as the distribution mean
for the parallel tempering method is already ill-defined for $L=6$.

\begin{table} \centering
  \caption{Ratios of the population mean $\tau_{\rm pop}$ and the quantile mean $\langle \tau \rangle_{\epsilon=10^{-4}}$  of the round-trip times for flat MUCA,
    the $1/k$-ensemble,
   and parallel tempering with respect to the same quantities for the power-law
   MUCA method.}

 \label{tab:relative_performance} \vspace{0.1cm}

 \begin{tabular}{ ccccccc }\hline
  \multicolumn{1}{c}{}& \multicolumn{2}{c}{flat
                        MUCA} & \multicolumn{2}{c} {$1/k$-ensemble} & \multicolumn{2}{c}{parallel tempering}\\\cline{1-7}
   $L$&$r_{\text{pop}}$&$r_{\epsilon=10^{-4}}$&$r_{\text{pop}}$&$r_{\epsilon=10^{-4}}$&$r_{\text{pop}}$&$r_{\epsilon=10^{-4}}$\\[1pt]\hline
& & & &\\[-9.5pt]

$3$ & $1.160(2)$ & $1.174(3)$& $1.0146(6)$ & $1.0193(9)$& $1.637(4)$ & $1.640(5)$\\
$4$ & $1.622(8)$ & $1.68(2)$& $1.288(5)$ & $1.328(10)$& $1.175(6)$ & $1.25(2)$\\
$5$ & $2.28(5)$ & $2.44(6)$& $1.63(3)$ & $1.75(4)$& $1.136(5)$ & $1.185(8)$\\
$6$ & $3.8(2)$ & $3.9(2)$& $2.59(9)$ & $2.6(2)$& $2.8(2)$ & $3.4(3)$\\
$8$ & $10.5(2)$ & $14.2(6)$& $6.9(3)$ & $9.4(4)$& $2.1(2)$ & $2.62(7)$\\
 \hline
    \end{tabular}
  \end{table}

 The two definitions are evaluated for all simulated lattice sizes and plotted in
Fig.~\ref{fig:means}.  Both definitions of the mean
grow exponentially up to linear system size $L=6$ until which the mean is defined,
while for $L>6$, where the distribution means diverge, they seem to be
growing faster than exponentially.  We have also looked at the scaling of
the more commonly used quantiles including the median~\cite{Berg2000,Bittner2006}, which
are derived directly from the $\tau$-values without the intermediate step of fitting
to a statistical model.
These quantiles behave similarly to the quantile means ~\eqref{eq:quantile_mean}
being, however, less stable for small $\epsilon$.

For the direct comparison of PL with the existing methods we introduce
the relative performance $r$ which we define as the fraction of the mean of the
respective method and the one of PL.  In
Table~\ref{tab:relative_performance}, the relative performance for all different
system sizes is listed.
The errors in $r$ are estimated using the Jackknife resampling technique.
It consists in generating a set of ratios $\{r_i\}$, where for the calculation
of each $r_i$ only a subset of all disorder realizations is taken.
The error in $r$ is derived from the variance of the so generated
Jackknife sample.

The speedup of PL compared to flat MUCA increases with
system size, reaching a factor of more than 10
for $L=8$ for both definitions of the mean, while compared to the
$1/k$ ensemble the speedup for the  biggest system size is still a factor of
$r\approx 7 - 9$. Compared to PT the
speedup is less pronounced and not steadily growing with system size,
reaching a factor of $r\approx 2-3$ for our largest system sizes.
\section{Conclusion}
\label{sec:conclusion}
Setting up multicanonical simulations such that the outcoming histograms are shaped
according to power laws instead of being flat is trivially achievable.
Nevertheless this simple
approach enables us to gather significantly more independent statistics at
the ground-state energy, which is important because the thermodynamic
contributions of the ground state of spin glasses are believed to be significant.
It is likely that similar techniques will also improve the sampling of the ground state
of other systems with complex free-energy landscape such as polymers and in
particular proteins, for
which the importance of the native state is well known.

While PT has been the most employed method in the simulation
of spin glasses probably also due to its good ability to investigate the
ground-state region, we were able to show that
the power-law setting considerably improves the performance of multicanonical
simulations in this respect,
rendering them at least comparable to PT.

The overall gain in performance grows with increasing lattice size and reaches a
factor of up to $10-15$ in comparison to flat MUCA and still a factor of up
to $2-3$ compared to PT.
In terms of round-trip time distributions the heaviness of the tails is
reduced by its superior ability to deal with the hard disorder realizations.

This improved ability of the here proposed power-law MUCA method of finding
ground states for the hard instances implies its usefulness in the
application to general optimization problems.  This is particularly useful
because many other optimization
problems can be rephrased in terms of spin-glass Hamiltonians~\cite{Lucas2014}
and thus solved employing the same methodology.

\section*{Acknowledgement}

This project was funded by the Deutsche Forschungsgemeinschaft (DFG, German Research Foundation) under project No.\ 189\,853\,844 -- SFB/TRR 102 (project B04). It was further supported by the Deutsch-Franz\"osische Hochschule (DFH-UFA) through the Doctoral College ``$\mathbb{L}^4$'' under Grant No.\ CDFA-02-07.

\end{document}